\documentclass[12pt,a4paper]{article}
\usepackage{amsfonts}
\usepackage{amssymb}
\usepackage{amsmath}

\usepackage{epsfig}
\usepackage{epstopdf}
\usepackage{graphicx}

\numberwithin{equation}{section}

\usepackage{cite}
\usepackage[colorlinks, linkcolor=blue,, urlcolor=blue, anchorcolor=blue, citecolor=blue]{hyperref}
\usepackage{breakurl}

\date{}

\begin{document}

\author{\small {Xue Zhang,$^a$
Ya-Bo Wu,$^a$\footnote{\href{mailto:ybwu61@163.com}{Corresponding author: ybwu61@163.com}}
~Song Li,$^b$ Yu-Chen Liu,$^a$ Bo-Hai Chen,$^a$ Yun-Tian Chai$^a$, Shuang Shu$^a$}}

\title{ Cosmological evolution of generalized non-local gravity}
\maketitle

{\small{\centerline{\small{$^a$ Department of Physics, Liaoning Normal University, Dalian 116029, China}}
{\small{\centerline{\small{$^b$ Department of Physics, Capital Normal University, Beijing 100048, China}}

\begin{abstract}
We construct a class of generalized non-local gravity (GNLG) model which is the modified theory of general relativity (GR)
obtained by adding a term $m^{2n-2} R\Box^{-n}R$ to the Einstein-Hilbert action.
Concretely, we not only study the gravitational equation for the GNLG model by introducing auxiliary scalar fields,
but also analyse the classical stability and examine the cosmological consequences of the model for different exponent $n$.
We find that the half of the scalar fields are always ghost-like
and the exponent $n$ must be taken even number for a stable GNLG model.
Meanwhile, the model spontaneously generates three dominant phases of the evolution of the universe,
and the equation of state parameters turn out to be phantom-like.
Furthermore, we clarify in another way that exponent $n$ should be even numbers by discuss the spherically symmetric static solutions in Newtonian gauge.
It is worth stressing that the results given by us can include ones in refs. \cite{arXiv:1402.0448,arXiv:1408.5058} as the special case of $n=2$.

\noindent{{\it Keywords:} modified gravity, dark energy theory}
\end{abstract}

\section{Introduction}
\label{sec:1}

In the last decade, it has been believed that our universe is undergoing an accelerating phase of expansion
which is indicated by the cosmological observations from the first Type Ia supernovae (SNe)\cite{astro-ph/9812133,astro-ph/9805201}
to the Wilkinson Microwave Anisotropy Probe (WMAP)\cite{arXiv:1212.5226} and Planck satellite experiment \cite{arXiv:1502.01589}.
Explaining the current phase of cosmic acceleration is an ongoing challenge.
The problem is that the geometry constructed from GR doesn't accord with the known matter from observation.
Different ways addressing this problem can be classified as the gravity side or the matter side of the Einstein equation.
On the one hand, the dark energy explains the data by introducing an exotic energy to the matter side
and provides the main contribution to the energy budget of the universe today.
However it arises a challenge to model builders who are attempting to understand the nature of it.
The simplest candidate for the dark energy is the cosmological constant $\Lambda$ \cite{astro-ph/9904398,astro-ph/0207347,astro-ph/0004075},
but the cosmological observations indicate that the dark energy may not be a constant.
In particular there is no good explanation for why the cosmological constant should be so small
and why it should recently have come into dominance.
On the other hand, the problem of understanding the origin of dark energy (DE) has stimulated a very active search
for modifications of General Relativity (GR) in recent years.
In particular, the recent observational constraints derived from Planck and other data show that
the dark energy equation of state $w_{de}$ smaller than $-1$ is favored \cite{arXiv:1502.01589,arXiv:1404.3467,U2}.
This may imply the infrared modification of gravity from general relativity,
because the models in the framework of GR, such as quintessence \cite{astro-ph/9807002,astro-ph/9908023,arXiv:1510.04010}
and k-essence \cite{astro-ph/9912463,astro-ph/0006373,arXiv:1503.03826},
generally predict $w_{de}$ larger than $-1$.
The challenge is to theoretically construct a consistent theory more effectively explaining the acceleration data,
and the significant deviations from GR are neglected by the data inside the solar system.
A typical modifications of GR is $f(R)$ model in which the Ricci scalar $R$ of the Einstein-Hilbert Lagrangian
replaced by an arbitrary function of the Ricci scalar
\cite{arXiv:0705.1158,arXiv:0705.3199,arXiv:0805.1726,arXiv:1002.4928,arXiv:1212.4921,arXiv:1404.1681}.
The $f(R)$ model represent the only local, metric-based, generally coordinate invariant and stable modification of gravity \cite{astro-ph/0601672}.
However, this model suffers from the major problem of an extra scalar degree of freedom.

More modification freedom is available if locality is abandoned,
a new class of non-local modifications of gravity has been introduced by Deser and Woodard \cite{arXiv:0706.2151}.
The model is constructed by adding to the Einstein-Hilbert action a term of $Rf(\Box^{-1}R)$.
The form of the non-local distortion function $f(\Box^{-1}R)$ can be chosen to reproduce the $\Lambda$CDM background cosmology exactly \cite{arXiv:0803.3399,arXiv:0904.0961}.
Absent a derivation from fundamental theory, the term $f(\Box^{-1}R)$ has the same status as the function $f(R)$ in $f(R)$ theory.
Moreover, the great advantage of this class of models is to trigger late time acceleration by the transition from radiation domination.
The Ricci scalar $R$ vanishes during radiation dominance
and $\Box^{-1}R$ cannot begin to grow until the onset of matter dominance
while its growth becomes logarithmic.
This theory have attracted some attention both theoretically and phenomenologically,
as a possible alternative to dark energy that presents the universe accelerated at late times.
However, this novel territory raises the worry of new degrees of freedom,
possibility of instability and negative energy \cite{arXiv:1311.3421}.
Furthermore, its cosmological perturbations has been studied in ref.~\cite{arXiv:1310.4329},
in which the Deser-Woodard model is ruled out by comparison with the data of structure formation.

Recently, Maggiore and Mancarella (MM) proposed a model based on the following non-local modification of General Relativity,
in which a mass parameter enters the theory as a suitable coefficient of the non-local term \cite{arXiv:1402.0448}
\begin{equation}
S_{MM}=\frac{1}{16\pi G}\int d^{d+1}x\sqrt{-g}\big[R-\frac{d-1}{4d}R\frac{m^2}{\Box^2}R\big],
\end{equation}
where $d$ is the spatial dimensions
and $(d-1)/4d$ is a normalized coefficient of the mass parameter $m$.
A natural way to proceed is to introduce a mass scale $m$ which is in order of the Hubble parameter' present value $H_0$.
Here the original term $f(\Box^{-1}R)$ is given by a specific form $m^2\Box^{-2}R$
which can be constrained by observations directly.
$\Box^{-1}$ operator is a formal inverse of $\Box$ in the scalar representation
which can be expressed as the convolution with a retarded Green's function $\mathcal{G}$ \cite{arXiv:1307.3898,arXiv:1112.4340,arXiv:1107.1463,arXiv:0910.4097,arXiv:0809.4927}.
In contrast with the non-local models of Deser-Woodard model,
the non-local term is controlled by a mass parameter $m$.
The model has the same number of parameters as one in $\Lambda$CDM with $\Omega_{\Lambda}$ replaced by $m$.
So no new scale is introduced in the theory.
Currently this model is receiving more attention \cite{arXiv:1408.5058,arXiv:1403.6068,arXiv:1408.1084},
because in contrast with other non-local models it can be constrained by observations directly \cite{arXiv:1311.3435,arXiv:1209.0836,arXiv:0807.3778,arXiv:1402.4613,arXiv:1307.6968,arXiv:1305.3034}.

Of course, if one try to explore the effect of non-local terms, there are many possible models that one could in principle study.
In this paper we will extend the MM model to the general form by adding a term $m^{2n-2} R\Box^{-n}R$ to the Einstein-Hilbert action,
which is here called the generalized non-local gravity (GNLG).
The meaning of such terms in the Lagrangian is obscure and not very well understood.
How to deal with this kind of Lagrangian is a non-trivial topic
and we will try to understand its content.
At the conceptual level, one might be worried by the presence of non-local terms in the equations of motion.
They emerge already in a purely classical context
from some form of iterative procedure in an underlying local fundamental theory.
So a range of possibilities exist for the construction of non-local models
and a number of general features can be extracted.
We try to understand the typical cosmological consequences of non-local terms that can be associated to a mass parameter $m$.
If this non-local model will turn out to fit the cosmological observations well,
this would provide a useful hint for the construction of the corresponding fundamental theory.
Hence, it is a natural problem that at the background level
whether there is a classical stability during the cosmological evolution
with features well compatible with the observations,
which is another aim in our paper.
We will perform a thorough stability analysis by taking into account all possible excitations of the metric to potentially drive an instability.
Throughout this paper we shall use the system of units in which $c=\hbar=1$
and the metric signature is $(-, +, +, +)$.

This paper is organized as follows.
In section~\ref{sec:2}, we construct a generalized non-local gravity (GNLG) model and give the related non-local formulation,
and we further study the stability of tensor modes in section~\ref{sec:3}.
In section~\ref{sec:4}, we analysis the cosmological evolution.
Spherically symmetric static solutions in Newtonian gauge are derived in section~\ref{sec:5}.
Finally, our conclusions and outlook are in section~\ref{sec:6}.

\section{generalized action and non-local formulation}
\label{sec:2}

We consider the generalized non-local gravity (GNLG) model defined by the action
\begin{eqnarray}\label{2.1}
S_{GNL}=\frac{1}{16\pi G}\int d^{d+1}x\sqrt{-g}R \Big(1-\lambda\frac{m^{2n-2}}{\Box^{n}}R \Big),
\end{eqnarray}
where the convenient normalization of the mass parameter $\lambda=(d-1)/4d$ and $d$ is the number of spatial dimensions.
The metric $g_{\mu\nu}$ is assumed to be minimally coupled to matter.
The generalized non-local Lagrangian has already been given.
It also remains to state is the covariant scalar d'Alembertian
\begin{equation}
\Box\equiv\frac{1}{\sqrt{-g}}\partial_{\mu}\big(\sqrt{-g}g^{\mu\nu}\partial_{\nu}\big).
\end{equation}
Its inverse means the retarded Green's function \cite{arXiv:1307.3898}.
This is in fact the standard way of proceeding
and it is the one used for the MM model.
Incidentally, both non-local terms $R\Box^{-2}R$,
or $(\Box^{-1}R)^2$ in the action will give the same equations of motion
and this effectively translates in the freedom of integrating by parts $\Box^{-1}$.
Now some issues need be clarified so that we can perform the latter variation. As discussed in refs.~\cite{arXiv:1308.2319,arXiv:1406.4998}
\begin{align}
&\delta(\Box)R=-\delta g_{\mu\nu}R^{;\mu;\nu}
+\frac{1}{2}g^{\mu\nu}R_{;\sigma}(\delta g_{\mu\nu})^{;\sigma}
-R^{;\mu}(\delta g_{\mu\nu})^{;\nu},\\
&R\delta(\Box^{-n})R=-\sum_{l=0}^{n-1}\Box^{l-n}R\delta(\Box)\Box^{-l-1}R.
\end{align}
The model is actually defined by its field equations,
which are obtained by varying the gravity and matter actions with respect to the metric.
This produces causal and conserved field equations.
Integrating by parts $\nabla$ and $\Box^{-1}$ a few times
we obtain the modified Einstein equations
\begin{equation}\label{2.5}
G_{\mu \nu}-\lambda m^{2n-2}K_{\mu \nu}=8\pi GT_{\mu \nu}
\end{equation}
with
\begin{align}\nonumber
K_{\mu \nu}=&\Big[2(G_{\mu \nu}-\nabla_{\mu}\nabla_{\nu}+g_{\mu \nu}\Box)+\frac{1}{2}g_{\mu \nu}R\Big](\Box^{-n}R)\\\nonumber
&+\sum^{n-1}_{l=0}\bigg\{\nabla_{\nu}(\Box^{l-n}R)\nabla_{\mu}(\Box^{-l-1}R)\\ \label{2.6}
&-\frac{1}{2}g_{\mu \nu}\Big[\nabla_{\sigma}(\Box^{l-n}R)\nabla^{\sigma}(\Box^{-l-1}R)+(\Box^{l-n}R)(\Box^{-l}R)\Big]\bigg\},
\end{align}
where $T_{\mu \nu}$ is the total energy momentum tensor of matter
and $K_{\mu \nu}$ comes from varying the non-local term in the above action.
The important property of this manipulation is that it preserves transversality $\nabla^{\mu}K_{\mu\nu}=0$.
This is the main advantage of starting with an action instead of modifying GR directly at the level of the equations of motion.
As a consequence, the energy-momentum tensor is conserved
and provide a sensible modification to classical GR.
Depending on the variational principle this additional prescription
makes the relation between the final equations of motion and the action rather formal.

We consider the model defined by the action \eqref{2.1} in $d=3$ spatial dimensions.
To generalize the definitions and localization procedure in ref. \cite{arXiv:1402.0448}
we introduce the following auxiliary scalar fields $U_1, \cdots, U_n$
such that we can rewrite the original GNLG model as a local form by a set of coupled differential equations
\begin{align}\label{2.7}
U_{1}=&-\Box^{-1}R,\\
U_{2}=&-\Box^{-1}U_{1}=\Box^{-2}R,\\
U_{3}=&-\Box^{-1}U_{2}=-\Box^{-3}R,\\
U_{4}=&-\Box^{-1}U_{3}=\Box^{-4}R,\\\nonumber
&\vdots\\\label{2.11}
U_{n}=&-\Box^{-1}U_{n-1}=(-1)^n\Box^{-n}R,
\end{align}
whose initial conditions are all zero.

Actually, to perform the numerical integration of these equations,
it can be more convenient to use a set of dimensionless variables $V_{i}= H_{0}^{2}U_{i}~(i=2, \cdots, n)$ and $h=H(t)/H_{0}$,
where $H(t)=\dot{a}/a$ and $H_{0}$ is the present value of the Hubble parameter.
We use $\ln{a}$ as the time coordinate and a prime denotes $\partial/\partial\ln a$.
From eqs. \eqref{2.7}-\eqref{2.11} we get
\begin{align}\label{2.12}
U_1''+U_1'(3+\zeta)&=6(2+\zeta),\\
V_2''+V_2'(3+\zeta)&=h^{-2}U_1,\\
V_3''+V_3'(3+\zeta)&=h^{-2}U_2,\\
V_4''+V_4'(3+\zeta)&=h^{-2}U_3,\\\nonumber
\vdots&\\\label{2.16}
V_n''+V_n'(3+\zeta)&=h^{-2}U_{n-1},
\end{align}
where $\zeta=h'/h$.

\section{Stability of tensor modes}
\label{sec:3}

In this section we follow the procedure in ref.~\cite{arXiv:1307.6639}.
We will see that the non-local equations require the same initial data,
subject to exactly the same constraints, as ones in GR.
Below, we will proceed for concreteness in a particular synchronous gauge
\begin{equation}
ds^2=-dt^2+h_{ij}dx^{i}dx^{j}.
\end{equation}
This is the synchronous time coordinate, namely, the proper time of free-falling observers,
which will become cosmic time when we go to consider cosmology.
The basic analysis and conclusions should be adapted to any gauge,
and here we will analyse the stability at linearized and kinematical level.

In synchronous gauge the covariant scalar d'Alembertian takes the form
\begin{equation}
\Box=-\partial_{0}^{2}-\frac{1}{2}h^{ij}\dot{h}_{ij}\partial_{0}+\frac{1}{\sqrt{h}}\partial_{i}(\sqrt{h}h^{ij}\partial_{j}).
\end{equation}
Here $h^{ij}$ denotes the inverse of the spatial metric $h_{ij}$,
$h$ stands for the determinant of $h_{ij}$ and an overdot represents differentiation with respect to time $t$.
we obtain the various curvature tensors and scalar
\begin{align}
R_{00}&=-\frac{1}{2}h^{kl}\ddot{h}_{kl}+\frac{1}{4}h^{ik}h^{jl}\dot{h}_{ij}\dot{h}_{kl},\\
R_{ij}&=\frac{1}{2}\ddot{h}_{ij}+\frac{1}{4}h^{kl}\dot{h}_{ij}\dot{h}_{kl}-\frac{1}{2}h^{kl}\dot{h}_{ik}\dot{h}_{jl}+\bar{R}_{ij},\\\label{R}
R&=h^{kl}\ddot{h}_{kl}+\frac{1}{4}h^{ij}h^{kl}\dot{h}_{ij}\dot{h}_{kl}-\frac{3}{4}h^{ik}h^{jl}\dot{h}_{ij}\dot{h}_{kl}+\bar{R},
\end{align}
where $\bar{R}_{ij}$ means the intrinsic spatial curvature as usual
\begin{equation*}
\bar{R}_{ij}=\Gamma^{k}_{ij, k}-\Gamma^{k}_{ik, j}+\Gamma^{k}_{ij}\Gamma^{l}_{kl}-\Gamma^{l}_{ikj}\Gamma^{k}_{jl}.
\end{equation*}

Furthermore, we rewrite the Ricci scalar as
\begin{equation}\label{rR}
R=-\Box \ln h+\mathcal{O}(\partial_0),
\end{equation}
where here $\Box$ is the d'Alembertian in the scalar representation.
Eqs. \eqref{R} \eqref{rR} show that $\Box^{-1}R$ contains only the fields $h_{ij}$ and their first-order derivatives of time.
We will check that the non-local corrections in eq.~\eqref{2.5} don't change the sign of the kinetic terms,
because of no ghost in GR.
Also one needs second-order derivatives of time (SODOT) on $\Box^{-1}R$ to obtain SODOT on $h_{ij}$.
Therefore, the kinetic terms only containing SODOT are either proportional to $G_{\mu\nu}$
or at least SODOT on $\Box^{-1}R$.
Obviously, the first three terms of non-local modified term $K_{\mu\nu}$ in eq.~\eqref{2.6} are only concerned.

The SODOT of $h_{ij}$ in the Einstein tensor are
\begin{equation}
G_{ij}=\frac{1}{2}\ddot{h}_{ij}-\frac{1}{2}h_{ij}\partial_0^2\ln h+\mathcal{O}(\partial_{0}), \quad h=\det h_{ij}.
\end{equation}
Note that only the first SODOT term is a kinetic term for the propagating modes
because $h$ is constrained by $G_{00}$ \cite{arXiv:1307.6639}.
At the same time, the second and third terms in eq.~\eqref{2.6} can be written as
\begin{equation}
(\nabla_i\nabla_j-h_{ij}\Box)\frac{1}{\Box^n}R=\nabla_i\nabla_j\frac{1}{\Box^n}R-h_{ij}\frac{1}{\Box^{n-1}}R,
\end{equation}
so that it actually contains no SODOT.
Therefore, the kinetic part of the non-local modified dynamical equation is
\begin{equation}\label{kpde}
\frac{1}{2}(1-2\lambda \frac{m^{2n-2}}{\Box^{n}}R)(\ddot{h}_{ij}-h_{ij}\partial_{0}^{2}\ln{h})+\mathcal{O}(\partial_{0})=8\pi GT_{ij}.
\end{equation}
We find that if $(1-2\lambda m^{2n-2}\Box^{-n}R)$ becomes negative, then the scalar fields become ghosts.
Certainly, we would like to discuss the cases of $(1-2\lambda m^{2n-2}\Box^{-n}R)>0$.
Here the mass term is chosen in the order of the Hubble scale today.
Therefore we study the quantities of background evolution on the FRW cosmology with radiation and matter in $d=3$.
To this end, using the definitions of eqs. \eqref{2.11} we find that
\begin{equation}
1-2\lambda \frac{m^{2n-2}}{\Box^{n}}R=1-3\gamma V_{n}, \quad \gamma\equiv\frac{2\lambda m^{2n-2}}{3H_0^2},
\end{equation}
where the dimensionless quantities $3\gamma V_n$ will be studied in next section.
We want to see whether the curves can cross the threshold value of $1$.
In other words, if the sign of the coefficient in front of the kinetic term in eq.~\eqref{kpde} changes in contrast to the one in GR,
tensor modes exhibit an instability, and we regard the scalar fields as ghosts.
As discussed in \cite{arXiv:1311.3421}, at the classical level a ghost will give rise to instability,
while at the quantum level it corresponds to a particle with negative energy.
Note that when taking $n=2$ the above equations all can reduce to the ones in ref.~\cite{arXiv:1408.5058}.

\section{Cosmological evolution}
\label{sec:4}

In this section, we study the cosmological consequences of the model at the level of background evolution.
We consider a flat FRW background in $d=3$ with metric
\begin{equation}
ds^2=-dt^2+a^2(t)d\mathbf{x}^2.
\end{equation}
From the (00) component of eq.~\eqref{2.6}, we get
\begin{equation}
K_{00}=(-1)^n\Big[6h^{2}V'_{n}+6h^{2}V_{n}-\frac{1}{2}h^{2}\Theta_1+\frac{1}{2}\Theta_2\Big],
\end{equation}
where
\begin{align}
\Theta_1&=\sum^{n-1}_{l=0}V_{n-l}'U_{l+1}',\\
\Theta_2&=\sum^{n-1}_{l=1}U_{n-l}U_l.
\end{align}
Moreover,
\begin{align}
\Theta_1'&=\sum^{n-1}_{l=0}\Big[-2(3+\zeta)V_{n-l}'U_{l+1}'-h^{-2}(U_{n-l-1}U_{l+1}'+U_{n-l}'U_{l})\Big],\\
\Theta_2'&=\sum^{n-1}_{l=1}(U_{n-l}'U_l+U_{n-l}U_l').
\end{align}

In this form, one finds that there is an effective dark energy density $\rho_{DE}=\rho_0\gamma Y$ where $\rho_0 =3H^2_0/(8\pi G)$,
$\gamma=m^2/(9H_0^2)$ and
\begin{equation}
Y=(-1)^n \Big(3h^{2}V'_{n}+3h^{2}V_{n}-\frac{1}{4}h^{2}\Theta_{1}+\frac{1}{4}\Theta_{2}\Big).
\end{equation}
Thus, we get
\begin{equation}\label{h2}
h^{2}=\frac{\Omega_M e^{-3x}+\Omega_R e^{-4x}+\frac{1}{4}\gamma(-1)^n \Theta_2}{1-\gamma(-1)^n(3V_n'+3V_n-\frac{1}{4}\Theta_1)}
\end{equation}
and
\begin{equation}
\zeta=\frac{h^{-2}(-3\Omega_M e^{-3x}-4\Omega_R e^{-4x})-3\gamma(-1)^n(-h^{-2}U_{n-1}+4V_n'-\frac{1}{2}\Theta_1)}{2\big[1-3\gamma(-1)^n V_n\big]},
\end{equation}
where $\Omega_M$, $\Omega_R$ are the fractional energy densities of matter and radiation, respectively.

Also, with $\zeta$ and $h^2$,
eqs. \eqref{2.12}-\eqref{2.16} provide a closed set of second-order differential equations for $V_1, \cdots, V_n$,
whose numerical integration is straightforward.
The results are illustrated in figure~\ref{fig:1}.
It is not difficult to see that these curves approach $1$ asymptotically for even number $n$,
which means that $(1-3\gamma V_n)>0$ at all times.
This proves the stability from the sign of the kinetic term of the GNLG model.
But for odd number $n$, the curves of term $3\gamma V_n$ are exponential growth with $\ln a$,
so the the coefficients in front of the kinetic term change sign during the cosmology evolution,
hence the scalar fields become ghosts in the future.

\begin{figure}[tbp]
\centering
\includegraphics[width=.495\textwidth, height=.35\textwidth]{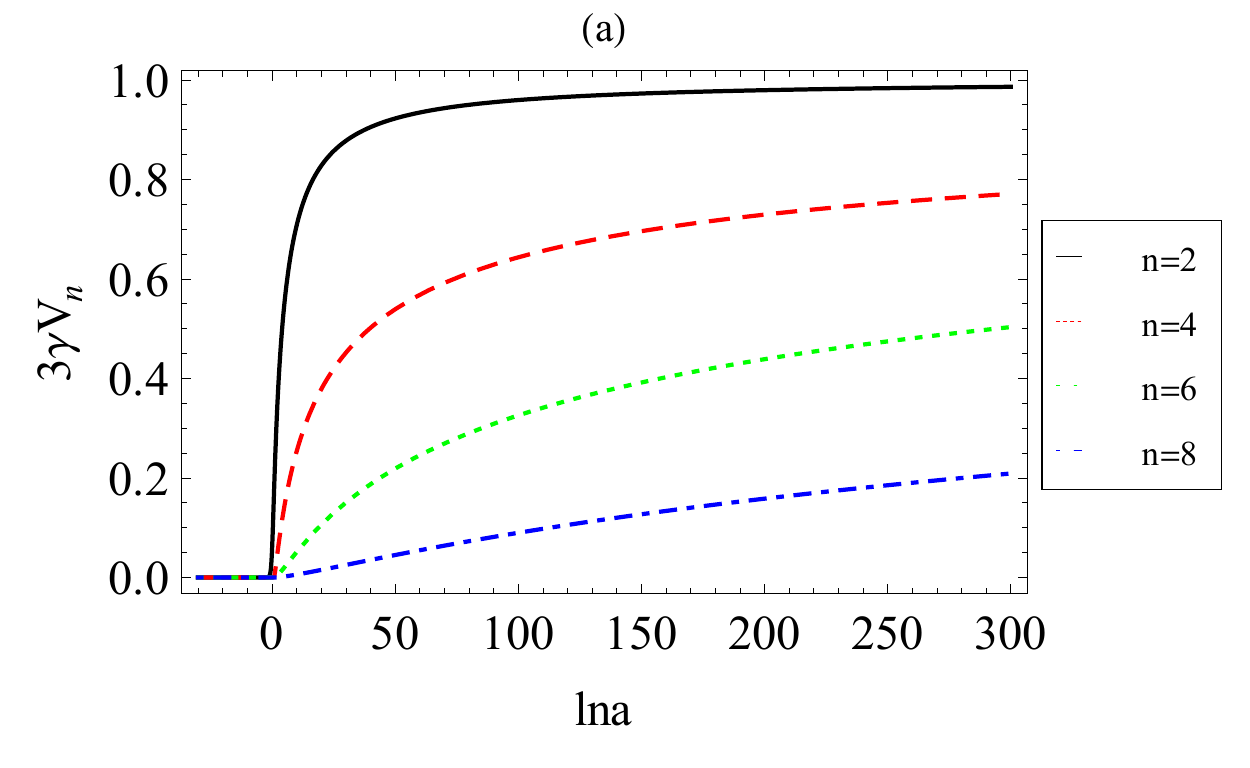}
\hfill
\includegraphics[width=.495\textwidth, height=.35\textwidth]{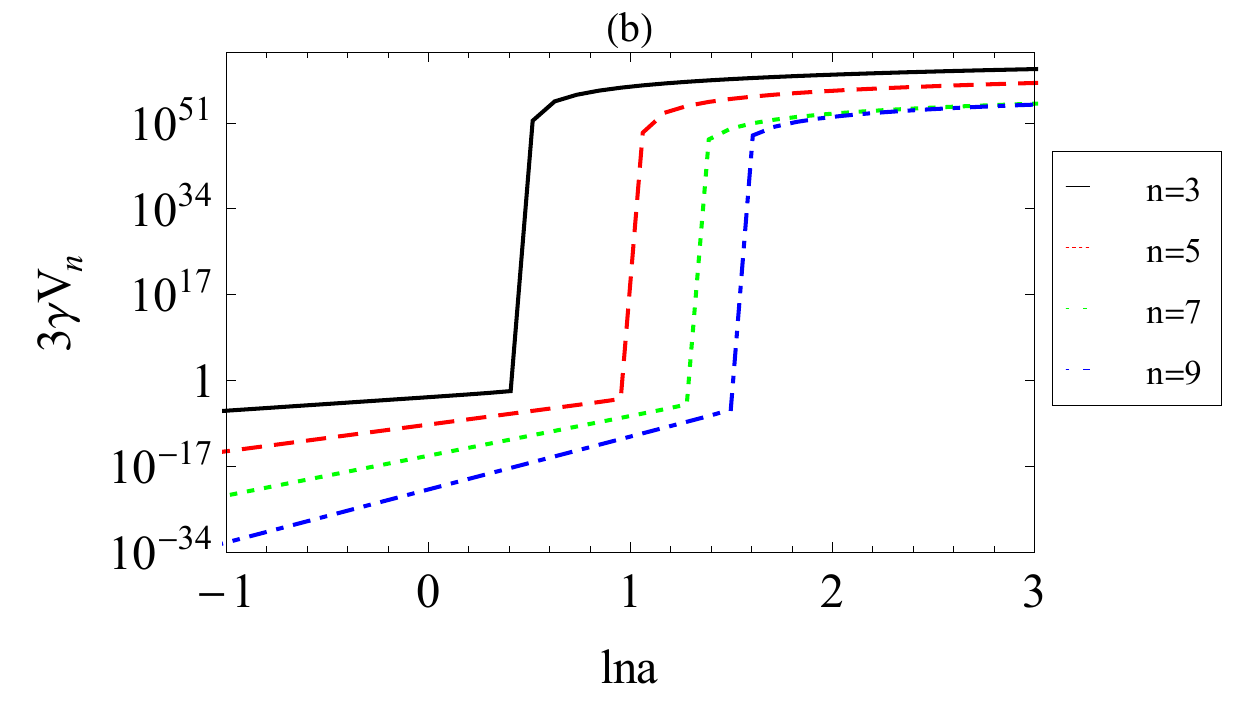}
\caption{\label{fig:1} \emph{Left panel:} the evolutions of $3\gamma V_n$ with $\ln a$ for even number $n$.
\emph{Right panel:} the evolutions of $3\gamma V_n$ with $\ln a$ for odd number $n$.}
\end{figure}

If we regard the non-local modification of space geometry as a component of dark energy,
from the conservation equation
\begin{equation}
\dot{\rho}_{DE}+3H(1+w_{DE})\rho_{DE}=0,
\end{equation}
we give the equation of state (EOS) for the dark energy
\begin{equation}
w_{DE}=-1+\frac{16h^{2}V_{n}'-8h^{2}\zeta V_n-4U_{n-1}-2h^{2}\Theta_{1}}
{12h^{2}V_{n}'+12h^{2}V_{n}-h^{2}\Theta_{1}+\Theta_{2}}.
\end{equation}
Concretely, we illustrate the evolutions of $w_{DE}$ for the even number $n$ in figure~\ref{fig:2}.
It is easy to see that the bigger $n$ is, the more obvious the perturbation of $w_{DE}$ in the past is.
The EOS $w_{DE}$ turn out to be on the phantom side, which is a general property of the non-local model,
and $w_{DE}<-1$ would generally imply a future singularity at a finite time \cite{astro-ph/0302506,astro-ph/0301273,astro-ph/9908168}.

\begin{figure}[tbp]
\centering
\includegraphics[width=.495\textwidth, height=.35\textwidth]{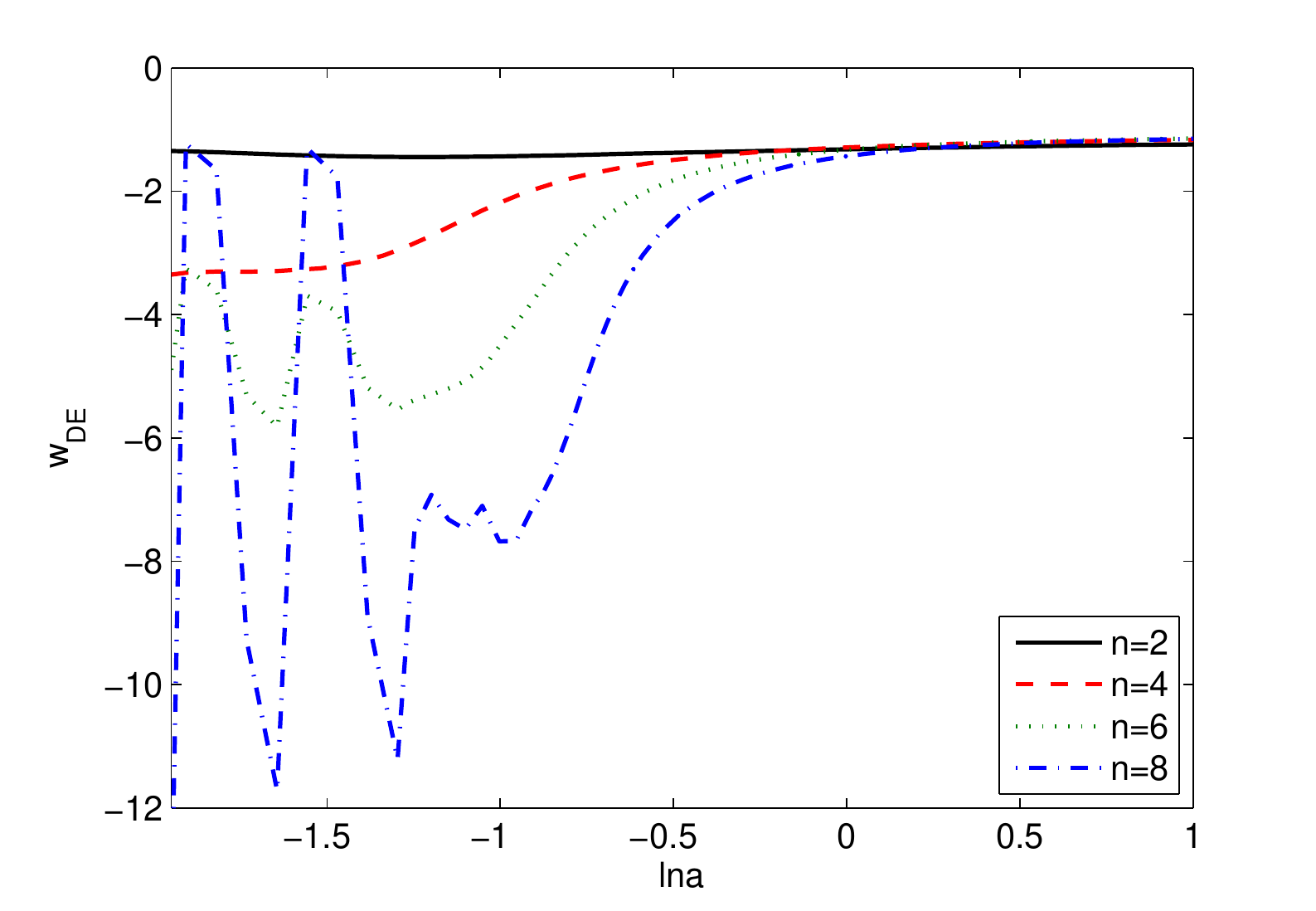}
\hfill
\includegraphics[width=.495\textwidth, height=.35\textwidth]{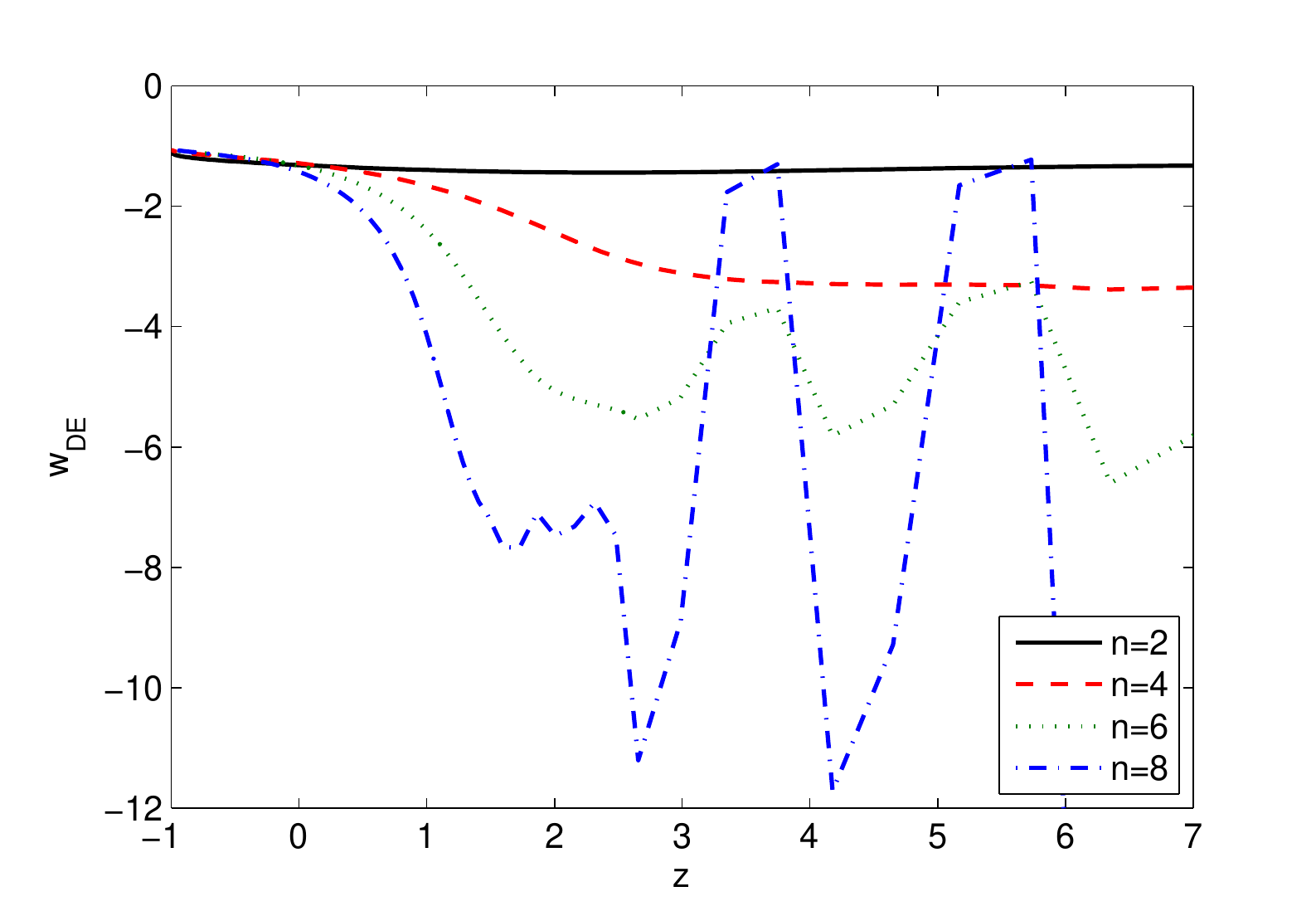}
\caption{\label{fig:2} \emph{Left panel:} the evolutions of $w_{DE}$ with $\ln a$ for even number $n$.
\emph{Right panel:} the evolutions of $w_{DE}$ with redshift $z$ for even number $n$.}
\end{figure}

Furthermore, we can rewrite the Friedmann equations in the usual form
\begin{align}
3H^2&=8\pi G\rho_{eff},\\
2\dot{H}+3H^2&=-8\pi G p_{eff},
\end{align}
which capture any deviation from GR as an effective fluid source, thus we get
\begin{equation}
w_{eff}=-1-\frac{2}{3}\zeta.
\end{equation}

\begin{figure}[tbp]
\centering
\includegraphics[width=.495\textwidth, height=.35\textwidth]{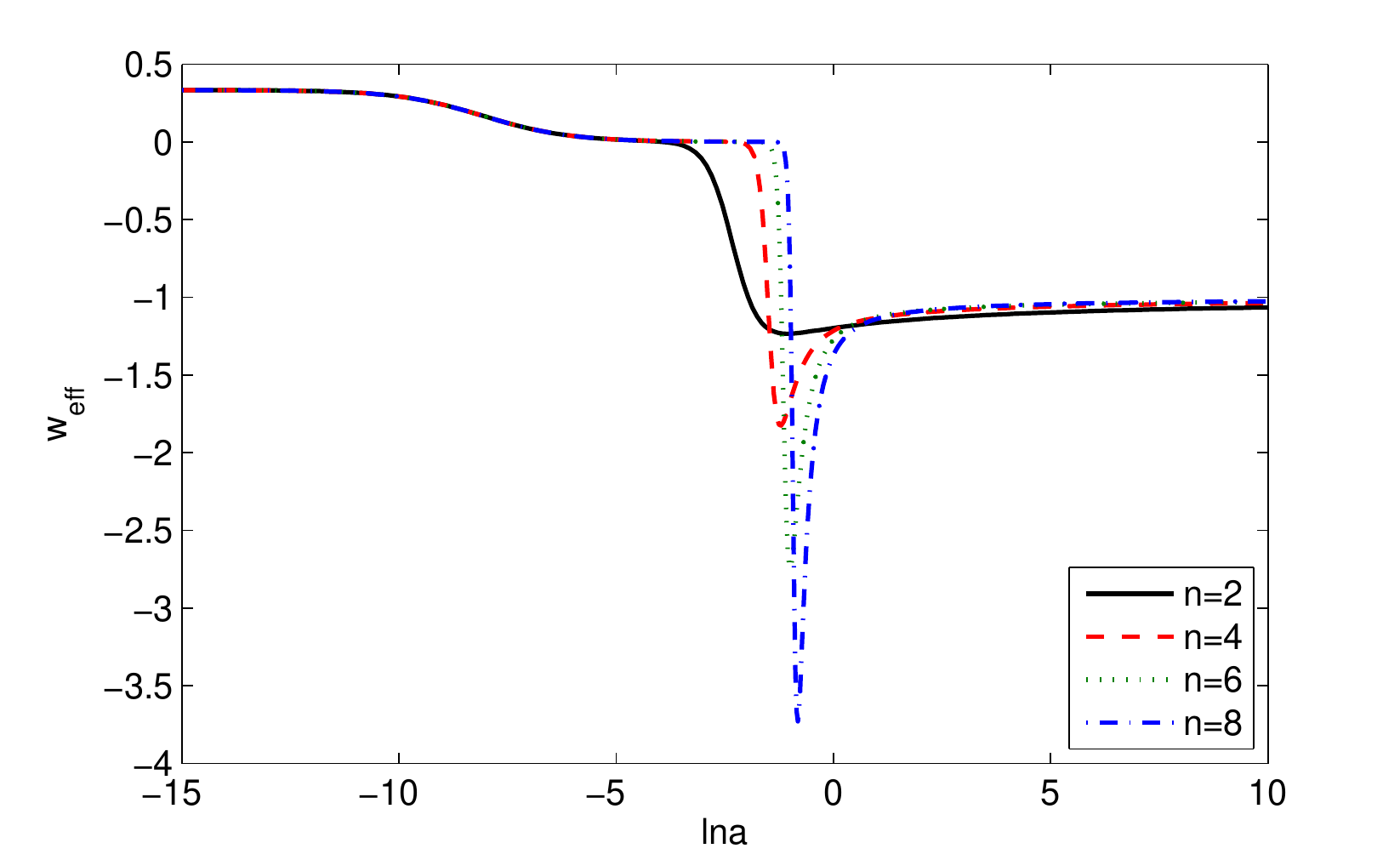}
\hfill
\includegraphics[width=.495\textwidth, height=.35\textwidth]{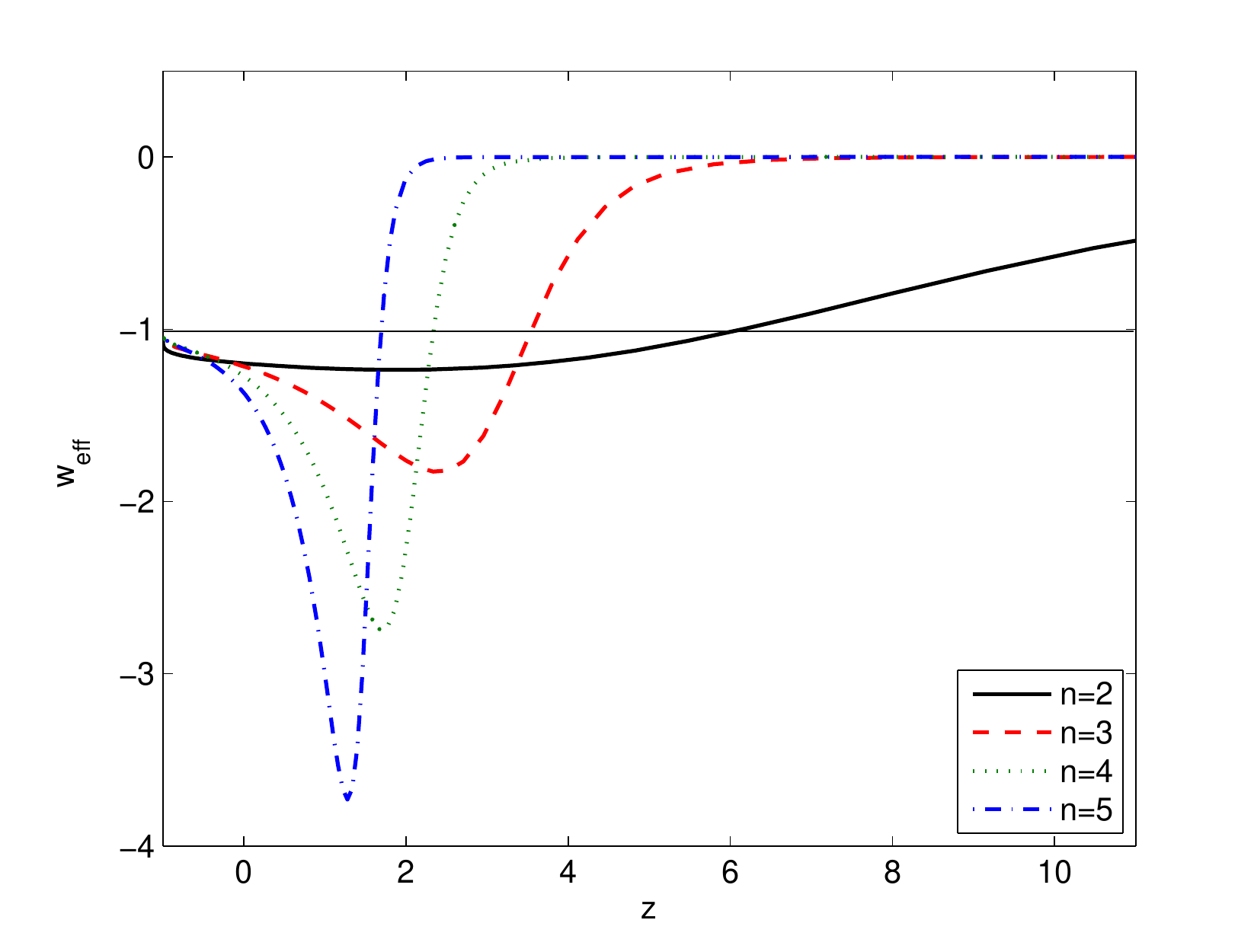}
\caption{\label{fig:3} \emph{Left panel:} the evolutions of $w_{eff}$ with $\ln a$ for even number $n$.
\emph{Right panel:} the evolutions of $w_{eff}$ with the redshift $z$ for even number $n$.}
\end{figure}

\begin{table}[tbp]
\centering
\begin{tabular}{|c|c|c|}
\hline
$n$ & $z_{min}$ & $w_{eff}(z=z_{min})$ \\
\hline
$2$ & $2.005$ & $-1.235$ \\
$4$ & $2.334$ & $-1.826$ \\
$6$ & $1.671$ & $-2.741$ \\
$8$ & $1.276$ & $-3.730$ \\
\hline
\end{tabular}
\caption{\label{tab:1}  The minimal value of $w_{eff}$ and the corresponding redshift $z_{min}$ for the different even number $n$.}
\end{table}

The evolutions of $w_{eff}$ with $\ln a$ and the redshift $z$ for even number $n$ are illustrated in figure~\ref{fig:3}, from which we find that the three flat values are $1/3$, $0$ and $-1$ corresponding to radiation dominant, matter dominant
and dark energy dominant phase, respectively.
Here we take $\Omega_{R}=0.0001$ and $\Omega_M=0.3175$ suggested by the Planck data assuming $\Lambda$CDM \cite{arXiv:1502.01589}.
We have tried various values of parameter $m$ and adjusted to the best value $m=10H_0$ finally.
The right panel of figure~\ref{fig:3} shows again the evolutional curves of $w_{eff}$ with the redshift $z$.
It is easy to see that the effect of DE vanishes in the deep radiation dominant phase
and then grows obvious during the matter dominant phase.
In addition, the EOS $w_{eff}$ of the effective fluid can cross the phantom divide $-1$,
and the bigger $n$ is, the earlier $w_{eff}$ crosses $-1$ in the past.
Meanwhile, there are the minimal values of $w_{eff}$ in the past not far from the present.
As shown in table 1, the corresponding minimal values of $w_{eff}$ to the different redshift $z_{min}$
in the four cases of $n=2, 4, 6$ and $8$ are $w_{eff}=-1.235, -1.826, -2.741$ and $-3.730$, respectively .
It follows that the bigger $n$ is, the smaller the minimal value of $w_{eff}$ is.
Note that when taking $n=2$ the above results all can reduce to the ones in ref.~\cite{arXiv:1402.0448,arXiv:1408.5058}.

\section{Spherically symmetric static solutions}
\label{sec:5}

We consider the effect of the source as a perturbation of Minkowski space $g_{\mu\nu}=\eta_{\mu\nu}+\delta g_{\mu\nu}$,
adapting the standard analysis performed in GR to recover the Newtonian limit \cite{arXiv:gr-qc/9712019},
namely, in the region $r_s\ll r$.
The most general static spherically symmetric (SSS) metric is
\begin{equation}
  ds^2=-e^{2\alpha(r)}dt^2+e^{2\beta(r)}dr^2+r^2(d\theta^2+\sin^2\theta d\phi^2).
\end{equation}
We focus on the scalar perturbations, so the perturbed metric can be rewritten as the form of Newtonian gauge \cite{arXiv:1401.8289}
\begin{equation}
  ds^2=-(1+2\Psi)dt^2+(1+2\Phi)dx^2.
\end{equation}
We also expand the auxiliary fields as $U_i=\bar{U}_i+\delta U_i (i=1, \cdots, n)$.
Keeping in mind that $\Psi$, $\Phi$ and $\delta U_i$ are first-order quantities.
If we consider a static source, partial derivatives on time vanish both for each background and the perturbation variable.
Therefore linearizing the $(00)$ and the $(ij)$ components of the modified Einstein equation eq.~\eqref{2.5}, we obtain
\begin{equation}\label{l00}
\nabla^2\Phi-\frac{m^{2n-2}(-1)^n}{6}\nabla^2\delta U_n=-4\pi G\delta\rho,
\end{equation}
\begin{equation}
\nabla^2(\Psi+\Phi)-\partial_i\partial_j(\Psi+\Phi)-\frac{m^{2n-2}(-1)^n}{3}(\nabla^2\delta U_n\delta_{ij}-\partial_i\partial_j\delta U_n)=0.
\end{equation}
Linearizing the equations of auxiliary fields $U_1, \cdots, U_n$ we get
\begin{align}\label{lU1}
&\nabla^2\delta U_1=\nabla^2(2\Psi+4\Phi),\\\label{lUn}
&\nabla^2\delta U_n=-\delta U_{n-1}.
\end{align}
Furthermore taking the trace of $(ij)$ components of the modified Einstein equations, we obtain
\begin{equation}\label{tlij}
\nabla^2\Big[\Psi+\Phi-\frac{m^{2n-2}(-1)^n}{3}\nabla^2\delta U_n\Big]=0.
\end{equation}
Note that the equations appeared in this section are only valid for the region $r\ll r_s$.
For a function $f$ satisfies $\nabla^2f=0$ at large $r$, one gets $f(r)=c_0+c_1r_s/r$ \cite{arXiv:1401.8289}.
In our case, the functions $\Psi$, $\Phi$ and $\delta U_i$ vanish at infinity, so the constant term $c_0$ is neglectful.
Then from eqs. \eqref{l00} \eqref{lU1} \eqref{lUn} \eqref{tlij} we get
\begin{equation}
\Big[\nabla^2+m^{2n-2}(-1)^n\Box^{2-n}\Big]\delta U_1=8\pi G\delta\rho.
\end{equation}
As discussed in ref. \cite{arXiv:gr-qc/9712019}, we can obtain the most general solution
\begin{equation}
\delta U_1=\frac{r_s}{r}\Big\{\cos[m^{n-1}(-1)^{n/2}\Box ^{1-n/2}r]+\beta\sin[m^{n-1}(-1)^{n/2}\Box ^{1-n/2}r]\Big\},
\end{equation}
\begin{align}
\Phi&=\frac{r_s}{2r}\Big\{c_{\Phi}+\frac{1}{3}\Big[\cos[m^{n-1}(-1)^{n/2}\Box ^{1-n/2}r]+\beta\sin[m^{n-1}(-1)^{n/2}\Box ^{1-n/2}r]\Big]\Big\},\\
\Psi&=\frac{r_s}{2r}\Big\{c_{\Psi}+\frac{1}{3}\Big[\cos[m^{n-1}(-1)^{n/2}\Box ^{1-n/2}r]+\beta\sin[m^{n-1}(-1)^{n/2}\Box ^{1-n/2}r]\Big]\Big\},
\end{align}
where $c_{\Phi}$, $c_{\Psi}$ and $\beta$ are arbitrary constants.
For simplicity in the expression, taking $A(r)=1+2\alpha(r)$ and $B(r)=1+2\beta$, namely, $A(r)=1+2\Psi$ and $B(r)=1-2r\Phi'$, we have
\begin{equation}
ds^2=-A(r)dt^2+B(r)dr^2+r^2(d\theta^2+\sin^2\theta d\phi^2),
\end{equation}
So
\begin{align}\nonumber
A(r)=&1+\frac{r_s}{r}\Big\{c_{\Psi}+\frac{1}{3}\Big[\cos\big(m^{n-1}(-1)^{n/2}\Box ^{1-n/2}r\big)\\
&+\beta\sin\big(m^{n-1}(-1)^{n/2}\Box ^{1-n/2}r\big)\Big]\Big\},\\\nonumber
B(r)=&1+\frac{r_s}{r}\Big\{c_{\Phi}+\frac{1}{3}\Big[\cos\big(m^{n-1}(-1)^{n/2}\Box ^{1-n/2}r\big)\\\nonumber
&+\beta\sin\big(m^{n-1}(-1)^{n/2}\Box ^{1-n/2}r\big)\Big]\\\nonumber
&+\frac{m^{n-1}(-1)^{n/2}\Box ^{1-n/2}r}{3}\Big[\sin\big(m^{n-1}(-1)^{n/2}\Box ^{1-n/2}r\big)\\
&-\beta\cos\big(m^{n-1}(-1)^{n/2}\Box ^{1-n/2}r\big)\Big]\Big\}.
\end{align}
In order to match these expressions with the GR solutions in the limit $m^{n-1}(-1)^{n/2}\Box ^{1-n/2}r\ll1$, we get $\beta=0$, $c_\Psi=-4/3$ and $c_\Phi=2/3$.
Furthermore we find that
\begin{align}
A(r)=&1-\frac{r_s}{r}\Big\{1+\frac{1}{3}\Big[1-\cos\big(m^{n-1}(-1)^{n/2}\Box ^{1-n/2}r\big)\Big]\Big\},\\\nonumber
B(r)=&1+\frac{r_s}{r}\Big\{1-\frac{1}{3}\Big[1-\cos\big(m^{n-1}(-1)^{n/2}\Box ^{1-n/2}r\big)\Big]\\
&+\frac{m^{n-1}(-1)^{n/2}\Box ^{1-n/2}r}{3}\Big[\sin\big(m^{n-1}(-1)^{n/2}\Box ^{1-n/2}r\big)\Big]\Big\},
\end{align}
while $\delta U_1=-\Box\delta^{-1}R$ is given by
\begin{equation}
\delta U_1=\frac{r_s}{r}\cos\big[m^{n-1}(-1)^{n/2}\Box ^{1-n/2}r\big].
\end{equation}
It follows that the exponent $n$ should here be the even number $n$, because of the appearance of the exponent $n/2$,
which is the same as the result given in section~\ref{sec:4}.

\section{Conclusion}
\label{sec:6}

We have constructed a class of generalized non-local gravity model
by adding a term $m^{2n-2}R\Box^{-n}R$ to the Einstein-Hilbert action.
Concretely, we not only studied this generalized non-local Lagrangian by introducing auxiliary scalar fields,
which evolve dynamically and do not correspond to extra degrees of freedom of the theory,
but also analysed the classical stability of GNLG model.
In figure~\ref{fig:1}, we illustrated the coefficient in front of the kinetic term to see whether its sign change or not.
The research results show that there is no ghost in the GNLG model with even number $n$,
but for odd number $n$, the scalar fields become ghost-like in the future.
In other words, the exponent $n$ must be taken even number for a stable GNLG model.
Moreover, we examined the cosmological evolutions of the GNLG model
and described the evolutional behaviors of the dark energy in figure~\ref{fig:2}.
We find that the dark energy turns out to be phantom-like,
and the bigger n is, the more obvious the perturbation of $w_{DE}$ in the past is.
The results show that the GNLG model spontaneously emerges three dominant phases of the evolution of the universe,
namely, radiation dominant, matter dominant and dark energy dominant phase, respectively.
Also, we find that the EOS $w_{eff}$ of the effective fluid can cross the phantom divide $-1$,
and the bigger $n$ is, the earlier $w_{eff}$ cross $-1$ in the past.
Meanwhile, from figure~\ref{fig:3} we know that there are the minimal values of $w_{eff}$ in the four cases of $n=2, 4, 6$ and $8$,
which are $w_{eff}=-1.235, -1.826, -2.741$ and $-3.730$, respectively.
It shows that the bigger $n$ is, the smaller the minimal value of $w_{eff}$ is.
Furthermore, we discussed the spherically symmetric static solution in Newtonian gauge,
and clarified in another way that the exponent $n$ should be even numbers.
It is worth stressing that the results given by us can include ones in refs. \cite{arXiv:1402.0448,arXiv:1408.5058} as the special case of $n=2$.

\section*{Acknowledgments}

We are deeply grateful to C. Cheng, K.C. Zhang, and Y.H. Xing for their helpful discussions and comments,
especially to Prof. Q.G. Huang for his directive help.
This work is supported by the National Natural Science Foundation of China (Grant Nos. 11175077, 11575075 and 11347163),
the Joint Specialized Research Fund for the Doctoral Program of Higher Education, Ministry of Education, China (Grant No. 20122136110002),
the Open Project Program of State Key Laboratory of Theoretical Physics, Institute of Theoretical Physics, Chinese Academy of Sciences, China (No. Y4KF101CJ1),
the Science and Technology Program Foundation of the Beijing Municipal Commission of Education of China under Grant No. KM201410028003
and the Project of Key Discipline of Theoretical Physics of Department of Education in Liaoning Province (Grant Nos. 905035 and 905061).

\end{document}